\documentclass{rspublicedit}

\usepackage{epsfig} 
\usepackage{amsmath}
\usepackage{amsbsy,color}

\newcommand{\beq}{\begin{equation}}
\newcommand{\eeq}{\end{equation}}
\newcommand{\beqa}{\begin{eqnarray}}
\newcommand{\eeqa}{\end{eqnarray}}
\newcommand{\ea}{{\it et al.\ }}
\newcommand{\om}{\Omega_m}

\newcommand{\gscr}{{\mathcal{G}}}
\newcommand{\vscr}{{\mathcal{V}}}

\begin{document} 

\title{Model Independent Tests of Cosmic Gravity} 
\author{Eric V.\ Linder}
\affiliation{Berkeley Center for Cosmological Physics \& Berkeley Lab, 
University of California, Berkeley, CA 94720, USA \\ 
Institute for the Early Universe, Ewha Womans University, Seoul, Korea} 

\date{\today}

\label{firstpage}

\maketitle 

\begin{abstract}{gravitation, general relativity, cosmic expansion, 
cosmic growth, galaxy surveys} 
Gravitation governs the expansion and fate of the universe, and the 
growth of large scale structure within it, but has 
not been tested in detail on these cosmic scales.  The observed 
acceleration of the expansion may provide signs of gravitational laws 
beyond general relativity.  Since the form of any such extension is 
not clear, from either theory or data, we adopt a model independent 
approach to parametrising deviations to the Einstein framework.  We 
explore the phase space dynamics of two key post-GR functions and 
derive a classification scheme and an absolute criterion on 
accuracy necessary for distinguishing classes of gravity models. 
Future surveys will be able to constrain the post-GR functions' 
amplitudes and forms to the required precision, and hence reveal 
new aspects of gravitation. 
\end{abstract}

\section{Introduction \label{sec:intro}}

Gravitation is the force that dominates the universe, from 
setting the overall expansion rate to forming the large scale 
structures of matter.  Yet our first precision tests of 
gravitation on large scales indicate our understanding of 
gravity acting on the components we see and expect -- baryonic 
and dark matter -- is insufficient.  The cosmic expansion is 
not decelerating as it should given these ingredients, but 
accelerating, pointing to either an exotic component with 
negative active gravitational mass (the sum of the energy 
density plus three times the pressure) or new aspects of 
gravity.  The growth and clustering of galaxies likewise do 
not agree with a universe possessing only gravitationally 
attractive matter within general relativity. 

Having been surprised in our first two tests of cosmic gravity, 
we naturally look to explore expansions of the classical 
framework.  Canonical general relativity (GR) with an exotic dark 
energy component, whether Einstein's cosmological constant or 
otherwise, is certainly one possibility.  Extending our knowledge 
of gravity is another, and is what this article concentrates on. 

One question is how to systematically analyse extensions to the 
known framework.  This can be done by working within an alternate, 
fully formed theory, but such first principles theories are scarce 
to nil (but see other articles within these Transactions for some 
possible guiding principles).  Instead we take a phenomenological, 
or (ideally) model independent, approach in analogy to the manner 
in which compact source gravity uses the parameterised post-Newtonian 
formalism.  Such model independence may put us in good stead to 
catch signs revealing the underlying nature of surprising observations. 

Building on a robust framework of the conservation and continuity 
equations, we parameterise functions from the equations of motion most 
closely tied to photon and matter density perturbation observables. 
This is equivalent to starting from the metric potentials themselves, 
although not directly from an action.  In Sec.~\ref{sec:par} we 
describe the parameterisation, and consistency structure of the 
system of equations, in more detail.  Discussing the types and reach 
of data enabled by future surveys such as BigBOSS, Euclid, and WFIRST, 
in Sec.~\ref{sec:constrain} we explore the degree of constraints that 
may be placed on these extended gravity, or post-GR, quantities in 
comparison to our knowledge today.  We draw connections between this 
model independent approach and representatives of strong coupling 
and dimensional reduction classes of gravity ($f(R)$ and DGP, 
respectively) in Sec.~\ref{sec:phase}, as well as exploring a different 
tack to distinguishing between gravity models through a phase space 
analysis similar to that used for dark energy.

\section{Framing Gravity \label{sec:par}}

In the equations of motion for cosmological perturbations in a 
homogeneous, isotropic background four quantities enter: the 
time-time and space-space metric potentials (equal to each other 
within GR), the mass density perturbation field, and the velocity 
perturbation field (taking a perfect fluid, e.g.\ ignoring 
pressure and anisotropic stress).  Conservation of stress-energy 
gives the continuity equation relating the density and velocity 
fields, and the Euler equation relating the velocity and time-time 
metric potential, so we are left with two free connecting equations.  
These can be chosen, for example, to be 
the gravitational slip between the two metric potentials and a 
modified Poisson equation relating the space-space potential to 
the density field, or they could be two Poisson-like 
equations relating one potential to the density field and the 
sum of the potentials to the density.  The latter choice turns out to 
give greater complementarity between the parameterised functions, 
with one closely tied to the growth of matter structures and the other 
closely related to photon perturbations such as gravitational lensing 
deflection and the integrated Sachs-Wolfe (ISW) effect. 

Table~\ref{tab:translate} shows several different forms of model 
independent parameterisation, and how they translate one into the 
other, adapted and extended from Daniel \ea 2010.

\begin{table}
\begin{tabular}{l l l}
Functions & Parameterisation & Reference\\ 
\hline
$\gscr$, $\vscr$ & Bins in $k$, $z$ & Daniel \& Linder 2010\\ 
$\mu=2\gscr-\vscr$, $\varpi=\frac{2\vscr-2\gscr}{2\gscr-\vscr}$ & 
$\mu,\varpi\sim a^s$ or bins in $z$ & See translation table in 
Daniel \ea 2010\\ 
PPF: $f_G=\gscr^{-1}-1$, $g=\frac{2\gscr}{2\gscr-\vscr}$ & model 
dependent & Hu \& Sawicki 2007, Hu 2008\\ 
MGCAMB: $\gamma=\frac{2\gscr}{\vscr}-1$, $\mu=\vscr$ & 
$\gamma,\,\mu=\frac{1+\beta_{\gamma,\mu}\lambda_{\gamma,\mu}^2 
k^2 a^s}{1+\lambda_{\gamma,\mu}^2 k^2 a^s}$ & Zhao \ea 2009, 
Bertschinger \& Zukin 2008\\ 
$\Sigma=\gscr$, $\mu=\vscr$ & $\Sigma=1+\Sigma_s a^s$, 
$\mu=1+\mu_s a^s$ & Song \ea 2010, Song \ea 2010b\\ 
& PCA & Zhao \ea 2010\\ 
\end{tabular}
\caption{Translation between several different parameterisations of 
extended gravity and the light/growth functions $\gscr$ and $\vscr$.
}
\label{tab:translate}
\end{table}

Recently, several groups (Song \ea 2010, Daniel \ea 2010, Zhao \ea 2010, 
Daniel \& Linder 2010) 
have advocated the ``light/growth'' forms we 
will use here, due to their close relations with observables and their 
near orthogonality.  The defining equations are 
\beqa 
-k^2(\phi+\psi)&=&8\pi G_N a^2\bar\rho_m\Delta_m \times \gscr \\ 
-k^2\psi&=&4\pi G_N  a^2\bar\rho_m\Delta_m \times \vscr\,. \label{eq:vscrdef} 
\eeqa 
where $\psi$ is the time-time metric potential, $\phi$ is the space-space 
metric potential (in conformal Newtonian gauge), $G_N$ is Newton's constant, 
$\bar\rho_m$ is the homogeneous matter density, $\Delta_m$ the perturbed 
matter density (gauge invariant), $k$ is the wavenumber, and $a$ is the 
scale factor.  The functions $\gscr(k,a)$ and $\vscr(k,a)$ generically are 
length scale and time dependent.  Arising from the sum of potentials, 
$\gscr$ (meant to evoke an effective Newton's constant) predominantly 
governs photon perturbations through light deflection and the ISW effect. 
Coming from the velocity equation, $\vscr$ predominantly governs growth 
and motion of structure.  They thus probe reasonably distinct areas of 
extension to standard gravity. 

These functions also map onto cosmological observations in somewhat 
orthogonal ways.  Cosmic microwave background (CMB) data is mostly 
sensitive to the ISW effect and 
hence $\gscr$ (since we want the gravitational modifications to be 
responsible for current acceleration, we assume at high redshift, e.g.\ 
at CMB last scattering, the theory acts like general relativity without 
modifications).  Weak gravitational lensing involves $\gscr$ through 
the light deflection law, and $\vscr$ to some extent through the growth 
of structure.  Galaxy distributions and motions are most 
sensitive to $\vscr$.  Note that the phenomenological gravitational 
growth index parameter $\gamma$ of Linder 2005, Linder \& Cahn 2007 is 
directly related to $\vscr$ (Daniel \ea 2010). 

Cosmological data, now or in the near future, will not have sufficient 
leverage to reconstruct the general functions $\gscr(k,a)$ and 
$\vscr(k,a)$.  Just as with the dark energy equation of state $w(a)$ 
describing the cosmic expansion, one can only constrain a very limited 
number of parameters describing the functions.  Some assume a 
particular time dependence, and possibly neglect scale dependence 
(e.g.\ Song \ea 2010b).  More generally, one can 
use principal component analysis to determine the best constrained 
eigenmodes of the functions (Zhao \ea 2009b).  
Here we will use a similar but simpler approach of dividing the 
functions into bins of redshift and wavenumber, 
since this provides a more direct physical interpretation of the 
results: gravity is modified in a certain way at low/high 
redshift and smaller/larger scales.  We find that two bins in redshift 
and two in wavenumber, for each of the two post-GR functions (we call 
this the $2\times2\times2$ gravity model), is the extent of the leverage 
that next generation surveys will provide, so more complex 
parameterisations are not useful.

\section{Constraining Gravity \label{sec:constrain}}

To effectively constrain the post-GR parameters of $\gscr(k_i,z_i)$ 
and $\vscr(k_i,z_i)$, where $i=1,2$ represent the two bins, we 
need observational data that is sensitive to both the effects on 
the photons (for $\gscr$) and the matter growth (for $\vscr$). 
Distance measures such as Type Ia supernova distances or baryon 
acoustic oscillation scales are useful for determining background 
quantities such as the matter density $\om$ that might have 
covariance with the post-GR parameters. 

For the current state of the art data we can consider CMB 
photon perturbation spectra (WMAP: Jarosik \ea 
2010), supernova distances (Amanullah \ea 2010), galaxy clustering 
(Reid \ea 2010), weak gravitational lensing (CFHTLS: Fu \ea 2008, 
COSMOS: Massey \ea 2007), and CMB temperature-galaxy 
crosscorrelation (Ho \ea 2008, Hirata \ea 2008).  The results, 
discussed in detail in Daniel \ea 2010, Daniel \& Linder 2010 
(also see Bean \& Tangmatitham 2010, and Lombriser \ea 2009, 
Thomas, Abdalla, \& Weller 2009 
for DGP constraints, Lombriser \ea 2010 for $f(R)$ constraints), 
show that while $\gscr$ 
is currently bounded to lie within 10-20\% of the GR value for 
each of the four combinations of low/high wavenumber and low/high 
redshift, $\vscr$ is only weakly limited to within $\pm1$ of the GR 
value.  This indicates that current growth probes do not have 
sufficient leverage to look for deviations from general relativity 
on cosmic scales.  Moreover, the two weak lensing data sets do not 
agree with each other, with COSMOS showing consistency with GR 
while CFHTLS gives up to 99\% cl deviations at high wavenumbers 
and low redshift.  This may be due to difficulties in extracting 
accurate weak lensing shear measurements on small angular scales 
where the density field is more nonlinear. 

To place significant limits on $\vscr$ and growth, future data sets 
including galaxy clustering and weak lensing surveys covering much 
more sky area, accurately, and ideally to greater depth are required. 
(Another, nearer term probe will be measurement of the CMB lensing 
deflection field.)  Surveys such as BigBOSS (Schlegel \ea 2009, 
Stril, Cahn, \& Linder 2010), 
Euclid (Refregier \ea 2010, Martinelli \ea 2010), KDUST (Zhao \ea 
2010b), LSST (LSST 2009), and WFIRST (Gehrels 2010) offer great 
potential gains.  Clear understanding of redshift space distortions 
would enable probes of the matter velocity field in addition to the 
density field and serve as a method to measure the growth rate (see 
below), giving further windows on extensions to gravitation theory. 

\begin{figure}[!t]
\begin{center} 
\psfig{file=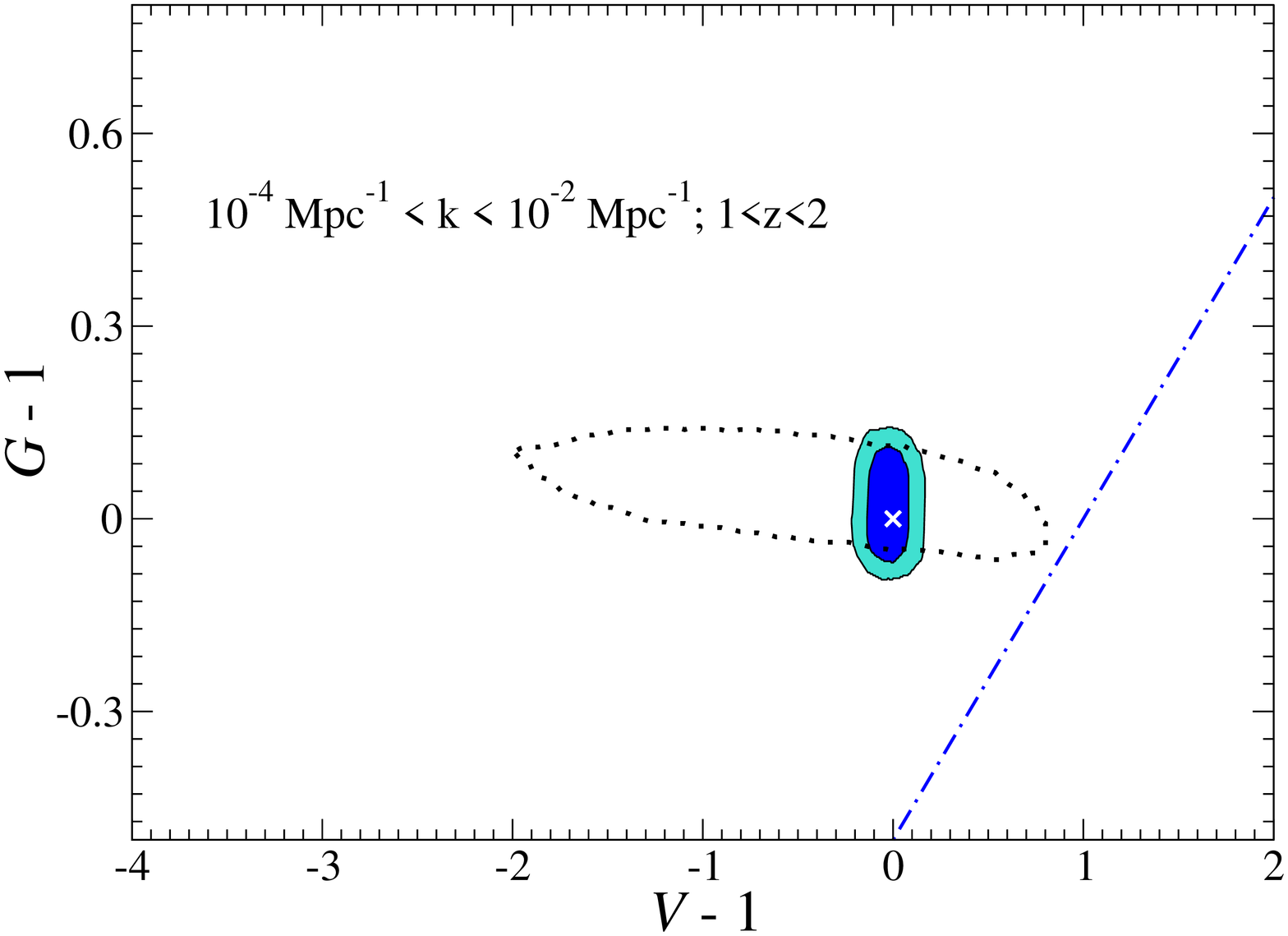,angle=-90,width=0.49\textwidth} 
\psfig{file=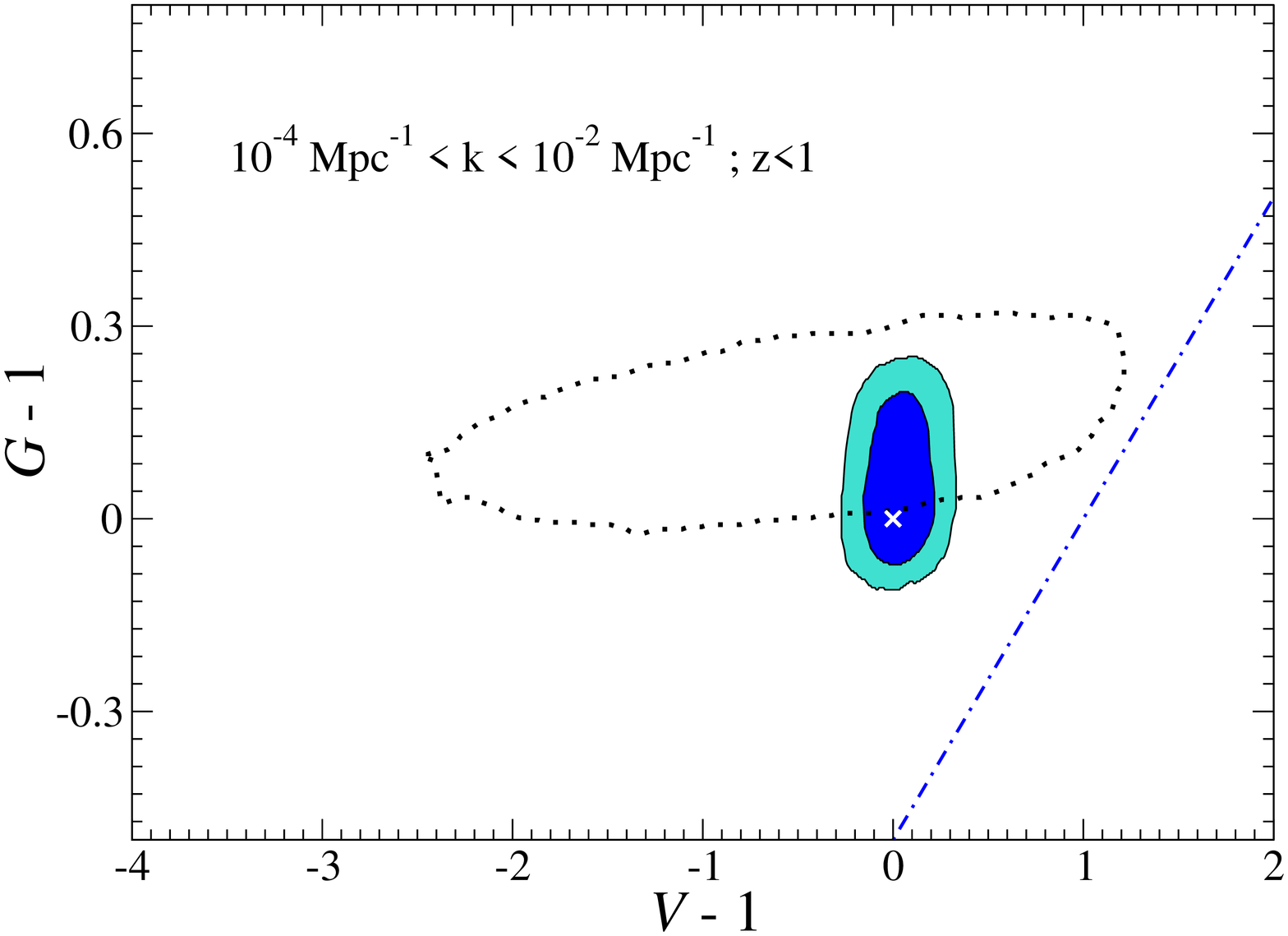,angle=-90,width=0.49\textwidth}\\ 
\psfig{file=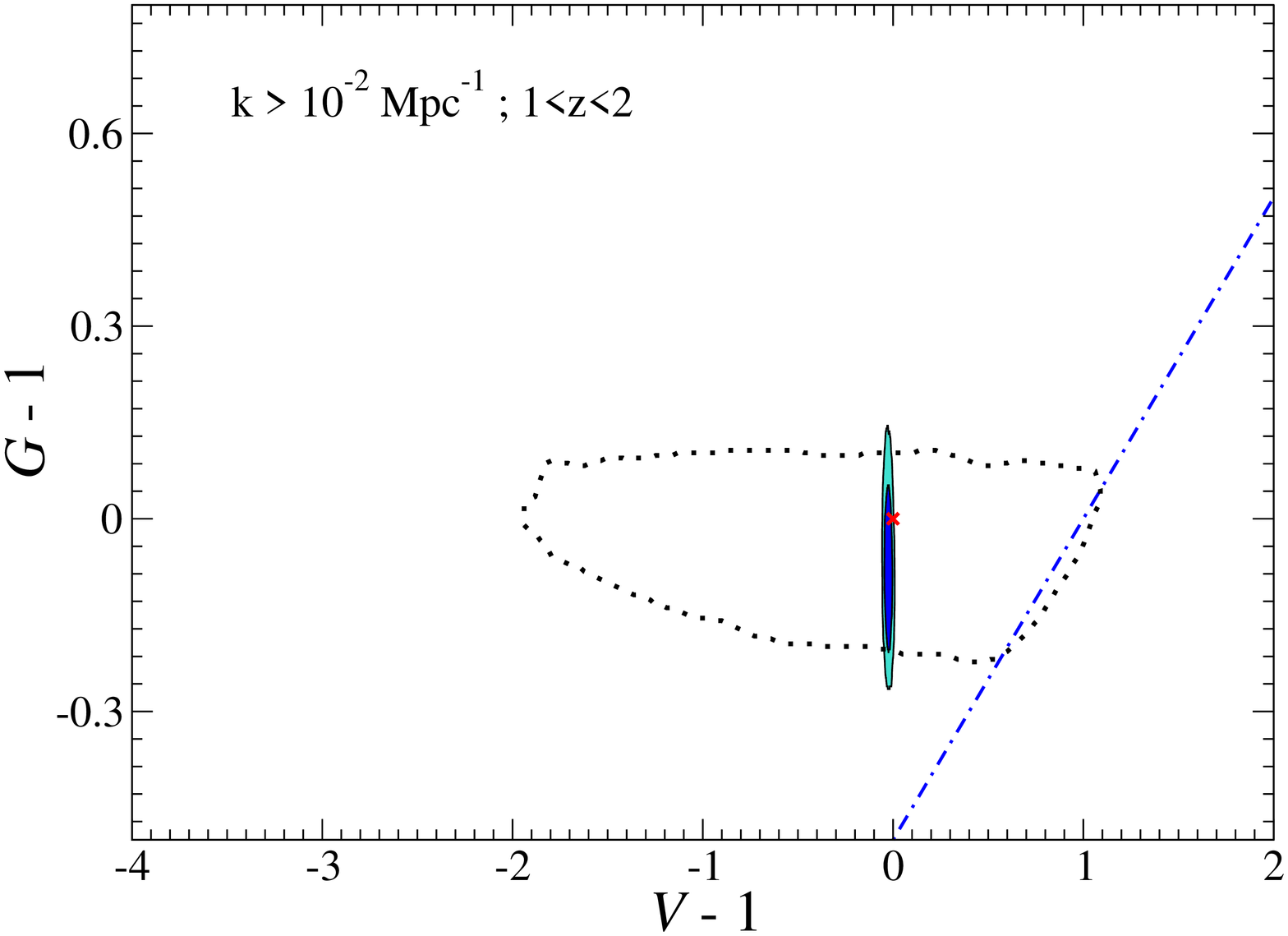,angle=-90,width=0.49\textwidth} 
\psfig{file=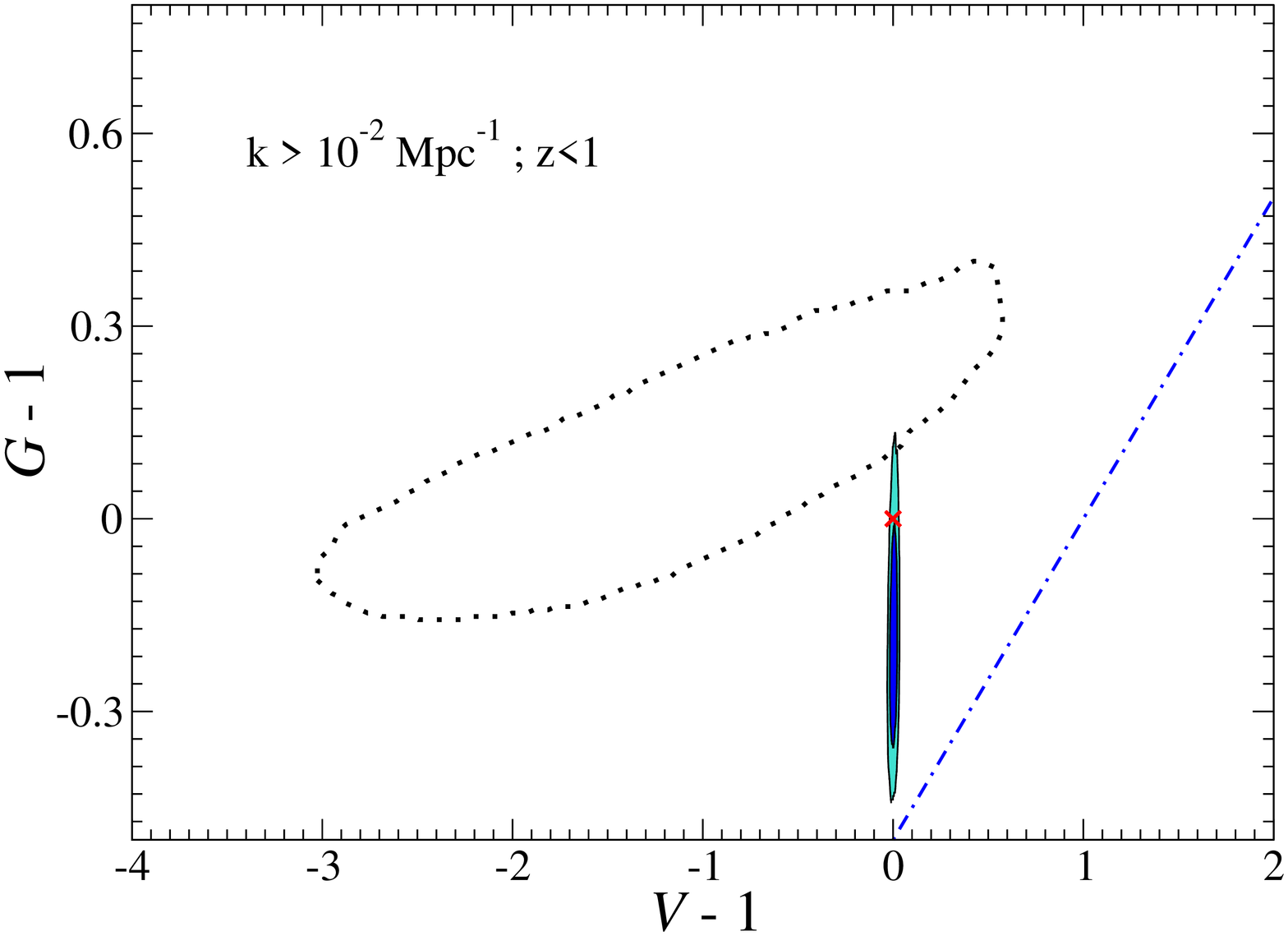,angle=-90,width=0.49\textwidth} 
\caption{
Filled contours show 68\% and 95\% cl constraints on $\vscr-1$ and 
$\gscr-1$ for the two redshift and two wavenumber bins using mock future 
BigBOSS, Planck, and WFIRST supernova data.  The dotted contours recreate 
the 95\% cl contours from Figs.~8 of Daniel \& Linder 2010 using current 
data, to show the expected improvement in constraints.  The x's denote 
the fiducial GR values (note the offset of current contours may be from 
systematics within the CFHTLS weak lensing data).  Adapted from 
Daniel \& Linder 2010.} 
\label{fig:bboss} 
  \end{center}
\end{figure}

Figure~\ref{fig:bboss} illustrates one example of future leverage 
possible on the $2\times2\times2$ post-GR model-independent 
parameterisation testing gravitation on cosmic scales.  The 
constraints on the growth and $\vscr$ tighten to the few--10\% 
precision level, giving tests of 8 different post-GR variables 
(and all their crosscorrelations) to better than 10\%.  
That could deliver strong guidance on the nature of cosmic gravity.

To compactify all the information on testing gravity into a single 
variable (which is not always desirable), 
we can also examine the gravitational growth index parameterisation 
approach and the constraints that future data will be able to place 
on $\gamma$.  Remember that this is only a partial characterisation 
of extensions to gravity, but can serve as an alert to deviations 
from general relativity (or to matter coupling) if the derived value 
of $\gamma$ shows time or scale dependence or is inconsistent with 
$\gamma_{GR}\approx0.55$. 

One promising method to measure $\gamma$ is redshift space 
distortions in the galaxy power spectrum (Linder 2008, Guzzo \ea 
2008).  This 
depends on the growth rate $f\equiv d\ln\Delta_m/d\ln a\approx 
\om(a)^\gamma$, as well as the growth itself $\Delta_m(a)$ and 
the galaxy bias $b(a)$.  One needs strong knowledge of the bias 
and modeling of the redshift distortion form (beyond linear theory), 
or excellent data (clear angular dependence maps or higher order 
correlations) to separate out $\gamma$ without assuming a form for 
the bias.  If the shape of the bias is fixed, keeping the amplitude 
as a fit parameter, then next generation galaxy surveys such as 
BigBOSS, Euclid, or WFIRST can measure $\gamma$ to $\sim7\%$, 
simultaneously with fitting the expansion history and neutrino mass 
effects on growth (Stril, Cahn, \& Linder 2010). 

Another probe is weak gravitational lensing, either by itself (in 
which case the mass is measured and galaxy bias does not enter) or 
in crosscorrelation with the galaxy density field (where one can form 
ratios of observables to separate out the galaxy bias).  Recall that 
extensions to gravity act on weak lensing in multiple ways: the growth 
of the matter power spectrum alters (which can be phrased in terms of 
$\gamma$), but also the light deflection law changes, involving the 
post-GR parameter $\gscr$ separate from $\gamma$.  This must also 
be included in the fit, except that many classes of extended gravity 
(such as DGP and $f(R)$ gravity) actually have $\gscr=1$ on 
cosmological scales.  One other subtlety is that the relation of the 
matter power spectrum to the photon temperature power spectrum changes 
with the altered growth, modifying the mapping between the primordial 
photon perturbation amplitude $A_s$ and the present mass amplitude 
$\sigma_8$.  When both weak lensing and cosmic microwave background 
data are included in the constraints, the treatment of $A_s$ and 
$\sigma_8$ must be made consistent. 

\begin{figure}
  \begin{center}{
  \includegraphics[width=\textwidth]{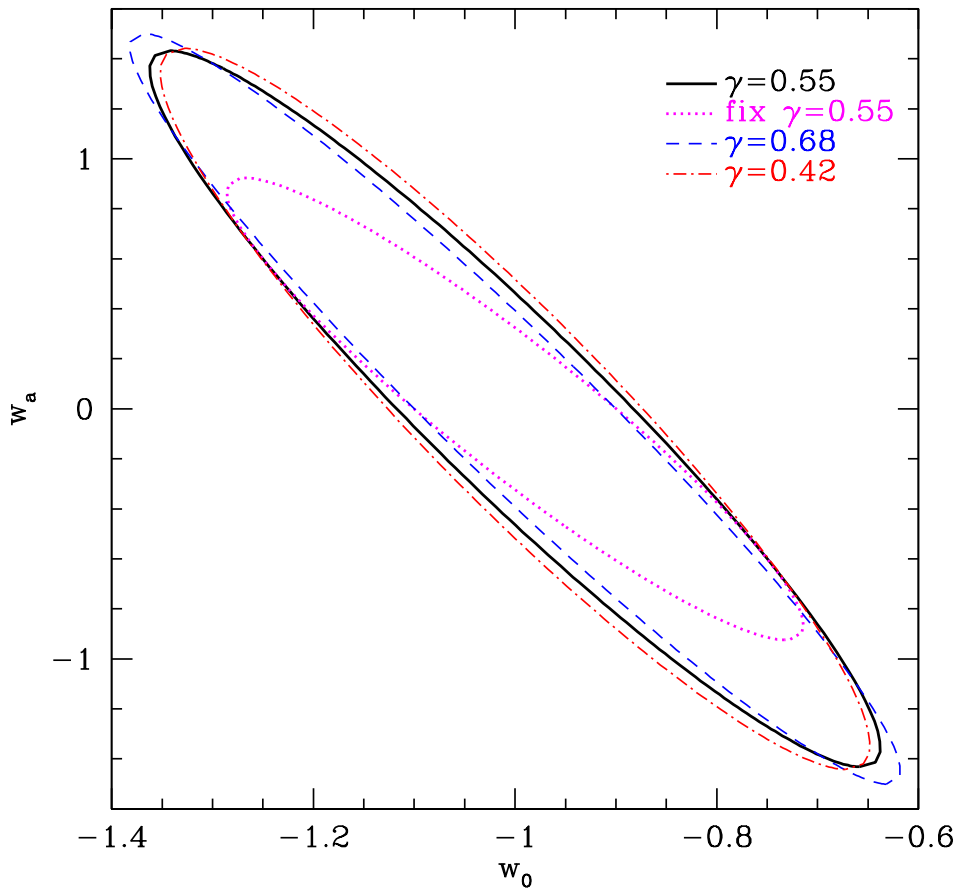}
  }
  \end{center}
  \caption{Weak gravitational lensing probes both the growth history 
and expansion history of the universe, so failure to account for 
possible extended gravity effects on the growth overestimates the 
tightness of constraints on the expansion history parameters $w_0$ 
and $w_a$.  Determination of $w_0$ and $w_a$ simultaneously with 
fitting for the growth index $\gamma$ is nearly independent of the 
value of $\gamma$, here shown with typical fiducial values for GR 
(0.55), DGP gravity (0.68), and $f(R)$ gravity (0.42).  The 68\% 
confidence level constraints on $\gamma$ are about 0.11 for the case 
shown of space-based weak gravitational lensing alone over 4000 
deg$^2$.  For the impact of other growth effects (e.g.\ neutrino 
mass, spatial curvature) on weak lensing see Das \ea 2011. 
}
  \label{fig:wlgamma} 
\end{figure}

Figure~\ref{fig:wlgamma} illustrates the effects of fitting the 
gravitational growth index $\gamma$ simultaneously with determining 
the effective dark energy equation of state $w(a)=w_0+w_a(1-a)$, 
i.e.\ the expansion 
history.  The area figure of merit in the $w_0$--$w_a$ plane 
decreases by 45\%, but as Huterer \& Linder 2007 pointed out the 
consequences of neglecting to fit for the growth index are worse.  
For an assumption mistaken by $\Delta\gamma$ and 
weak lensing alone as a probe the derived value of $w_a$ would be 
biased as $\Delta w_a\approx 8\Delta\gamma$.  When fitting for 
$\gamma$ simultaneously, however, the estimation of $w_0$ and $w_a$ 
do not depend strongly on the fiducial $\gamma$.  This is a 
reflection of $\gamma$ being defined specifically to separate the 
expansion history influence on growth from any  ``beyond general 
relativity'' effects (Linder 2005), making it a key method to test 
gravity.

\section{Paths of Gravity \label{sec:phase}}

While the model independent approach allows exploration without 
assuming a particular theory, it is of interest as well to consider 
some specific models and their mapping into the post-GR parameters 
we have described.  One can talk about three broad classes of 
extensions to gravity in terms of the physics restoring them to 
general relativity in solar system conditions, as observations require 
(Jain \& Khoury 2010, also see Durrer 2011, Maartens 2011, Uzan 2011 
in this volume): 
dimensional reduction where below a Vainshtein radius the theory 
acts like GR (e.g.\ DGP or cascading gravity), strong coupling 
where extra degrees of freedom freeze out through gaining a large 
mass in a chameleon mechanism (e.g.\ $f(R)$ or scalar-tensor gravity), 
or screening where the extra degrees of freedom decouple and vanish 
through symmetry restoration (symmetron gravity).  On cosmic scales, 
the first and third classes behave similarly, so we examine two 
representative cases: DGP gravity and $f(R)$ gravity. 

In both of these cases, the light variable $\gscr$ is simply equal 
to unity, the GR value.  However the growth variable $\vscr$ is 
affected.  The expressions become 
\beqa 
\vscr_{DGP}&=&\frac{2+4\om^2(a)}{3+3\om^2(a)} \label{eq:vdgp}\\ 
\vscr_{f(R)}&=&\frac{3+4\kappa^2(k,a)}{3+3\kappa^2(k,a)} 
\,, \label{eq:vfr}
\eeqa 
where $\om(a)=8\pi G_N\bar\rho_m(a)/[3H^2(a)]$ is the dimensionless 
matter density, $H=\dot a/a$ is the Hubble expansion rate, and 
$\kappa=k/[aM(a)]$ where $M(a)\approx (3\,d^2f/dR^2)^{-1/2}$ is the 
effective scalar field 
mass.  Note that DGP gravity does not have scale dependence on cosmic 
scales above the Vainshtein scale; gravity is scale free (e.g.\ the 
force is a power law with distance) on both the large scale 
(5-dimensional gravity) and small scale (4-dimensional gravity) 
limits and only the Vainshtein scale defined from the 5-d to 4-d 
crossover breaks this.  On the other hand, $f(R)$ gravity has 
both scale and time dependence, though tied together in a specific 
manner. 

Recall that the gravitational growth index $\gamma$ is related 
to $\vscr$.  For DGP gravity, $\gamma=0.68$ (Lue, Scoccimarro, \& 
Starkman 2004, Linder 2005, Linder \& Cahn 2007) 
is an excellent approximation to use for calculating the matter density 
linear growth factor as a function of redshift, good to 0.2\%.  
A mild time dependence can be incorporated into $\gamma$ (though 
this is not necessary) through Eq.~(27) of Linder \& Cahn 2007.  
For $f(R)$ gravity, $\gamma$ is not generally as well approximated by 
a constant in time, and has non-negligible scale dependence at 
redshifts $z\approx1$--3 (e.g.\ Tsujikawa \ea 2009, Motohashi, 
Starobinsky, \& Yokoyama 2010, Appleby \& Weller 2010) 
although 
the details depend on the specific $f(R)$ model and parameters. 

From Eqs.~(\ref{eq:vdgp})-(\ref{eq:vfr}) we can create phase plane 
diagrams of the evolution of the post-GR function $\vscr$.  This 
is analogous to the dark energy phase plane $w$--$w'$ for the dark 
energy equation of state and its time variation, where prime denotes 
$d/d\ln a$.  For dark energy, such diagrams led to clear distinction 
of certain physical classes (Caldwell \& Linder 2005) as well as 
calibration of the compact and accurate parameterisation 
$w(a)=w_0+w_a(1-a)$ (de Putter \& Linder 2008).  (Also see 
Song \ea 2010 for comparison of $\gscr$ and $\vscr$ at a fixed time 
in extended gravity vs interacting dark energy.) 

\begin{figure}
  \begin{center}{
  \includegraphics[width=\textwidth]{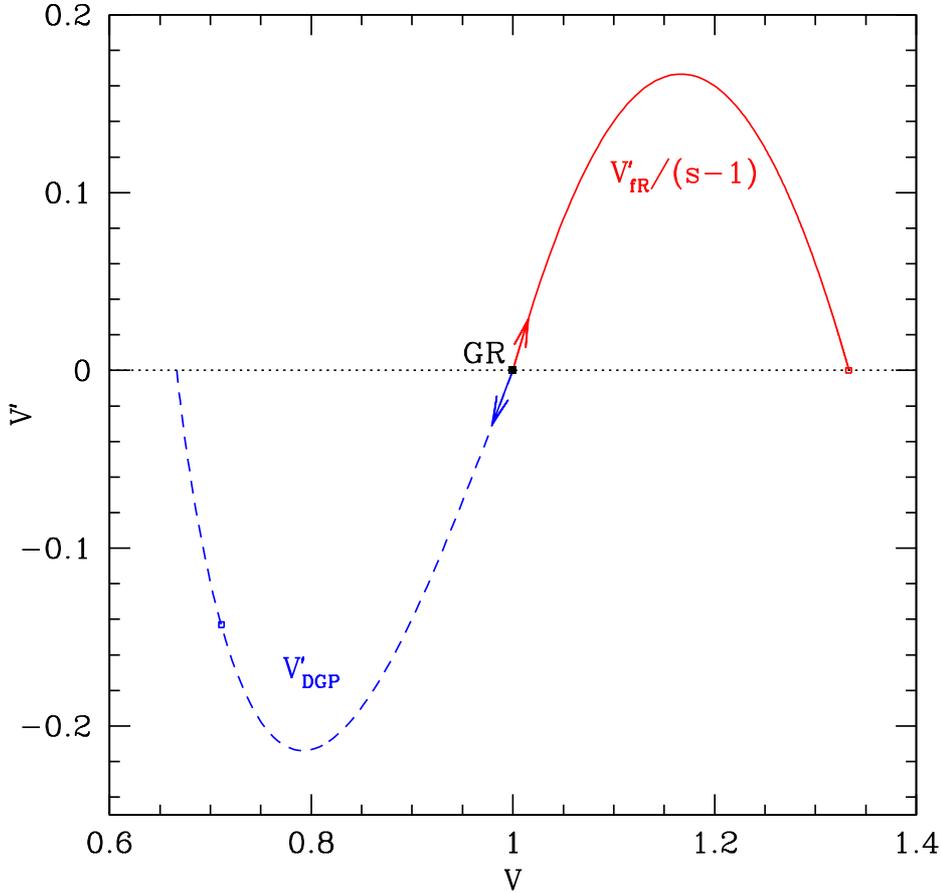}
  }
  \end{center}
  \caption{The phase space trajectories of DGP gravity and 
$f(R)$ gravity resemble thawing dark energy, except here the 
theories move away from general relativity in the post-GR 
growth parameter plane, instead of thawing from a cosmological 
constant.  Unlike canonical dark energy, the theories thaw 
in opposite directions: DGP moves to weaker gravity than GR 
while $f(R)$ moves to stronger gravity.  The phase space 
location today for each of the theories is shown by the squares 
(for $\om=0.27$). 
} 
  \label{fig:frdgp}
\end{figure}

Figure~\ref{fig:frdgp} shows the results in the 
$\vscr$--$\vscr'$ plane for DGP and $f(R)$ gravity.  For DGP gravity 
the equation for the phase space trajectory is 
\beqa 
\vscr'&=&\frac{-4\om^2(1-\om)}{(1+\om)(1+\om^2)^2} \\ 
&=&-18\,(1-\vscr)\left(\vscr-\frac{2}{3}\right) 
\left[1+\sqrt{\frac{3\vscr-2}{4-3\vscr}}\right]^2 \,, 
\eeqa 
where we used the relations 
\beq 
\om'=3w\om(1-\om)=-3\om\frac{1-\om}{1+\om} \,, 
\eeq 
and $\om$ is the time dependent matter density. 

For $f(R)$ gravity, the phase space trajectory is 
\beqa 
\vscr'&=&\frac{2\kappa\kappa'}{3(1+\kappa^2)^2} \\ 
&\to&\frac{2(s-1)\kappa^2}{3(1+\kappa^2)^2}= 
-6(s-1)(\vscr-1)(\vscr-4/3) \,, 
\eeqa 
where in the second line we parameterise $M(a)=M_0 a^{-s}$. 
If we wanted to look at the phase space evolution with 
respect to  inverse length $k$ rather than time $a$, then 
the equation still holds, with $s=2$.

Both classes of gravity theory act as thawing cases in the 
nomenclature of dark energy phase space: they are frozen 
in the general relativity state $(\vscr,\vscr')=(1,0)$ in the 
past, then as the Hubble parameter or Ricci scalar curvature 
drop from the cosmic expansion the theory thaws and moves away 
from GR.  The theories eventually freeze to an asymptotic 
attractor with $\vscr'=0$ in the future, with $\vscr=2/3$ in 
the case of DGP gravity (weaker gravity) and $\vscr=4/3$ in 
the case of $f(R)$ gravity (stronger gravity).  Note that 
interpreting $f(R)$ as a scalar-tensor theory agrees with the 
expectation that gravity should strengthen, since forces carried 
by a scalar field are attractive.  The clear separation of 
phase space territory for the two classes, as for thawing and 
freezing fields of dark energy, shows how observations could 
distinguish the nature of gravity.  This also defines a science 
requirement that $\vscr$ should be measurable to an accuracy 
better than 0.1 for a $3\sigma$ distinction of gravity theories. 

However, there is one further point regarding $f(R)$.  While the 
form $M(a)\sim a^{-s}$ is a reasonable description for the past 
behavior of the scalar field mass (see, e.g., Bertschinger \& 
Zukin 2008, Zhao \ea 2009, Appleby \& Weller 2010), 
in the future we expect $M$ to freeze to a constant (e.g.\ as $R$ 
itself does when the theory goes to the de Sitter attractor state; 
thanks to Stephen Appleby for pointing this out). 
If instead we parameterise the scalar field mass as 
$M(a)=M_1a^{-s}+M_*$ (accurate to $\sim1\%$ for at least some 
models), then the phase space trajectory does 
not asymptote to $(4/3,0)$ but rather returns to the GR limit of 
$(1,0)$ as $a\gg1$ and $\kappa=k/(aM)\to0$ in the future. 

\begin{figure}
  \begin{center}{
  \includegraphics[width=\textwidth]{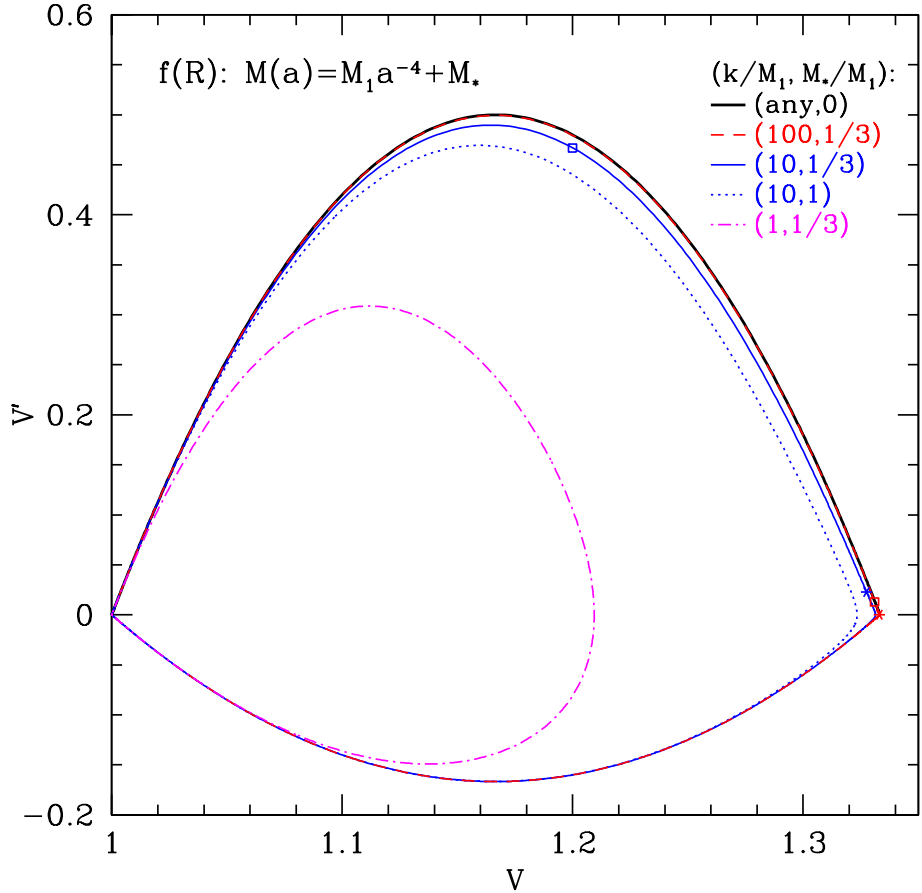}
  }
  \end{center}
  \caption{In the 
$f(R)$ case with a de Sitter future the scalar field mass $M$
freezes at a non-zero constant, causing the theory to restore
GR on cosmic scales just like with the chameleon mechanism on
small scales.  The heavy black curve recreates the $M_*=0$,
vanishing mass case of Fig.~\ref{fig:frdgp} that freezes at
$\vscr=4/3$, however the finite mass cases return to $\vscr=1$.
While the wavenumber dependence of the post-GR parameter at
$z=1$ (squares) is appreciable, by $z=0$ the $k$-dependence
is negligible (stars).  Since the gravitational growth index
$\gamma$ is directly related to $\vscr$, this explains the
scale independence of $\gamma$ at the present in such $f(R)$ 
models. 
}
  \label{fig:frfut}
\end{figure}

Figure~\ref{fig:frfut} illustrates this behavior.  As $M_*$ 
increases (dotted curve) or for wavenumbers near the mass scale 
$M$ (dot-dashed curve), the cosmic version of the chameleon 
mechanism begins to operate and the trajectory heads back toward 
GR.  Also note that because $M$ is 
no longer a power law in $a$, different wavenumbers $k$ 
do not simply correspond to a rescaling of $a$ and so the single 
$f(R)$ trajectory in Fig.~\ref{fig:frdgp} breaks up into varied 
paths for different $k$.  The evolution along a path varies as 
well.  This can be 
seen by the different values of $\vscr$ at $a=0.5$ for $k/M_1=10$ 
(blue square near the peak of the figure) vs $k/M_1=100$ (red 
square near the right side).  However, by the present (and for 
several e-folds to the future) this $k$ 
dependence has vanished, with $\vscr(a=1)$ shown with stars for the 
two cases agreeing to within 0.5\%.  This holds for all $k/M_1\gg1$ 
(note that the scalar mass $M_1$ is likely to be of order the 
Hubble constant for observationally allowed models).  Only for 
$k/M_1\approx1$ does scale dependence today enter (and the 
trajectory as a whole deviate).  Since the gravitational growth 
index $\gamma$ is directly related to $\vscr$, this explains the 
scale independence of $\gamma$ in such $f(R)$ models at the
present (see, e.g., Tsujikawa \ea 2009, Appleby \& Weller 2010).

\section{Conclusions} 

Cosmic acceleration may be a sign that gravitation deviates from 
general relativity on large scales, pointing the way to a deeper 
theory of gravity and perhaps the nature of spacetime and fundamental 
physics.  Even apart from this, cosmological observations now have 
the capability to test gravity on scales barely probed, and we 
should certainly do so. 

In the absence of a compelling specific model, and to remain 
receptive to surprises in how gravity behaves, a model independent 
approach in terms of parameterising the relations between the 
metric potentials and matter density and velocity observables 
has advantages.  We have explored here the ``$2\times2\times2$'' 
approach parameterising light/growth functions $\gscr$, $\vscr$ 
motivated by observable effects on photon and matter perturbations.  
These are divided into bins sensitive to both scale and time 
dependence. 

We find strong complementarity between $\gscr$ and $\vscr$, with 
each probing specific aspects of extensions to gravity, and both 
capable of being constrained by a variety of cosmological methods. 
Current surveys are making some inroads on determining $\gscr$, 
but constraining $\vscr$ requires future large surveys such as 
BigBOSS, Euclid, or WFIRST.  Crosscorrelations between different 
probes will be valuable as well, and geometric measures such as 
supernova and BAO distances will be essential to fit simultaneously 
the cosmic expansion history.  Conversely, fitting extended 
gravity growth parameters such as $\gamma$ when using probes such 
as weak lensing to measure the equation of state is necessary to 
avoid bias. 

Viewing the post-GR parameters in a dynamical phase plane yields 
insights similar to its use for dark energy.  Models such as 
DGP and $f(R)$ gravity act as thawing fields, although evolving 
in opposite directions away from GR.  Phase diagrams also can 
illustrate the scale dependence of $\gamma$ at various epochs, 
and most importantly deliver a science requirement on distinguishing 
classes of gravity: future surveys should aim to determine $\vscr$ 
to better than 0.1.  Planned next generation experiments such as 
BigBOSS, Euclid, and WFIRST can indeed potentially reach this level, 
with the caveat that expansion history should be tested as well, 
such as through supernova distances immune to gravitational 
modifications. 

Gravity can and should be tested on all scales, on laboratory, 
solar system, compact object, galactic, cluster, cosmological, and 
horizon scales.  The approach discussed here is designed for model 
independence on cosmological scales, smaller than the horizon. 
On horizon scales the light/growth functions become more complicated 
or insufficient as other terms in the equations become important 
(see, e.g., Bertschinger \& Zukin 2008, Hu \& Sawicki 2007, Ferreira 
\& Skordis 2010). 

Another model independent approach is to check 
consistency relations within general relativity (see, e.g., Zhang \ea 
2007, Acquaviva \ea 2008, Reyes \ea 2010).  These can provide an 
alert to deviations, and then one must adopt specific models or 
more detailed parameterisations such as discussed here to characterise 
the physics.

\section*{Acknowledgments}

I gratefully acknowledge Stephen Appleby, Scott Daniel, and Tristan Smith 
for valuable discussions and collaborations.  I thank the Centro de Ciencias 
Pedro Pascual in Benasque, Spain and the Kavli Royal Society International 
Centre for hospitality. 
This work has been supported in part by the Director, Office of Science, 
Office of High Energy Physics, of the U.S.\ Department of Energy under 
Contract No.\ DE-AC02-05CH11231, and the World Class University grant 
R32-2009-000-10130-0 through the National Research Foundation, Ministry 
of Education, Science and Technology of Korea.


\label{lastpage}

\end{document}